\documentclass[prl,twocolumn,showpacs,superscriptaddress,preprintnumbers,amsmath,amssymb,floatfix]{revtex4}
\usepackage{bm}
\usepackage{graphicx}
\usepackage{amsbsy}
\usepackage{amsmath}
\usepackage{amsfonts}
\usepackage{amsthm}
\usepackage{color}
\usepackage{mathrsfs}
\usepackage{graphicx}
\usepackage{dcolumn}
\usepackage{bm}

\newcommand{\bwt}{\begin{widetext}}
\newcommand{\ewt}{\end{widetext}}
\newcommand{\bea}{\begin{eqnarray}}
\newcommand{\eea}{\end{eqnarray}}

\begin{document}

\title{Efficient multi-mode  quantum memory based on photon echo in optimal QED cavity}

\author{Sergey A. Moiseev}
\email{samoi@yandex.ru}

\affiliation{Kazan Physical-Technical Institute of the Russian Academy of Sciences,
10/7 Sibirsky Trakt, Kazan, 420029, Russia}
\affiliation{Institute for Informatics of Tatarstan Academy of Sciences, 20 Mushtary, Kazan, 420012, Russia}

\affiliation{Physical Department of Kazan State University, Kremlevskaya 18, Kazan, 420008, Russia}.

\author{Sergey N. Andrianov}
\affiliation{Institute for Informatics of Tatarstan Academy of Sciences, 20 Mushtary, Kazan, 420012, Russia}

\affiliation{Physical Department of Kazan State University, Kremlevskaya 18, Kazan, 420008, Russia}.

\author{Firdus F. Gubaidullin}

\affiliation{Kazan Physical-Technical Institute of the Russian Academy of Sciences,
10/7 Sibirsky Trakt, Kazan, 420029, Russia}

\affiliation{Institute for Informatics of Tatarstan Academy of Sciences, 20 Mushtary, Kazan, 420012, Russia}

\date{\today}

\begin{abstract}

Effective multi-mode photon echo based quantum memory on multi-atomic ensemble
in the QED cavity is proposed. Analytical solution is obtained for the quantum memory efficiency that can be equal unity when optimal
relations for the cavity and atomic parameters are held. Numerical estimation for realistic atomic and cavity parameters demonstrates the high efficiency of the quantum memory for optically thin resonant atomic system.

\end{abstract}



\maketitle

Quantum communications and quantum computation require an effective quantum
memory (QM) that should possess a multi-mode and high fidelity character \cite{Briegel1998, Nielsen2000, Kok2007,Lvovsky2009,Tittel2010}. Most well-known QM based on electromagnetically induced transparency effect \cite{Fleischhauer2000} demonstrates an  efficient  storage  and retrieval only for a specific single temporal mode regime \cite{Gorshkov2007,Novikova2007, Appel2008}. Photon echo QM \cite{Moiseev2001,Moiseev2003,Kraus2006,Alexander2006,Gouet2009} offers most promising properties for realization of the multi-mode QM \cite{Simon2007, Nunn2008,Gisin2010}. However, the quantum efficiency of all discussed multi-mode variants of the  photon echo QM tends to unity for infinite optical depth $\alpha L$ as $[1-exp(-\alpha L)]^2$, where $\alpha$ and $L$ are resonant absorption coefficient and length of the medium along the light field propagation \cite{Moiseev2004,Sangouard2007}. It imposes a fundamental limit for the QM efficiency so it is necessary  to increase either the atomic concentration or the medium length. However, the QM device should be compact and the large increase of the atomic concentration gives rise to atomic decoherence due to the dipole-dipole interactions limiting thereby a storage time. So, using the free space QM scheme is quite problematic for practical devices. Efficient photon echo QM with controlled reverse of inhomogeneous broadening (CRIB) have been studied recently  in ideal cavity \cite{Moiseev2006} and in bad cavity \cite{Gorshkov2008} where high QM efficiency has been demonstrated only for a specific optimal single mode regime.
Here, we propose a general approach for multi-mode photon echo type of QM in QED cavity (single mode resonator). We demonstrate a high efficiency of the QM  for the optimized system of atoms and QED cavity at arbitrary  temporal shape of the stored field modes. We find a simple analytical solution for QM efficiency and the optimal conditions for matching of the atomic and cavity parameters where the QM efficiency can reach unity  even for small optical depth of the medium loaded in the cavity.

\emph{Basic equations:}
We analyze resonant multi-atomic system in a single mode QED cavity coupled with signal and bath fields.
By following to the cavity mode formalism \cite{Milburn}, we use a Tavis-Cumming Hamiltonian \cite{Tavis}
$\hat {H} = \hat {H}_o + \hat {H}_1 $,
for N atoms, field modes and their interactions taking into account the inhomogeneous and homogeneous broadenings of the atomic frequencies and continuous spectral distribution of the field modes where

\begin{align}
\label{eq1}
\hat {H}_o = &
\hbar \omega _o \{{\sum\nolimits_{j =1}^{N } {\hat {S}_z^{j} } }
+ \hat {a}^ + \hat {a}
\nonumber   \\&
+\sum\nolimits_{l =
1}^2 {\int {\hat {b}_l^ + (\omega )\hat {b}_l (\omega )d\omega } } \},
\end{align}

\noindent
are main energies of atoms ($S_{z}^j $ is a z-projection of the spin $\textstyle{1/2}$
 operator),
energy of cavity field ($\hat {a}^ + $ and $\hat {a}$ are
arising and decreasing operators), energies of signal (l=1) and
bath (l=2) fields ($b_l^ + $ and $b_l $ are arising and
decreasing operators of the field modes $[\hat {b}_l^ + (\omega '),\hat
{b}_{l'} (\omega )] = \delta _{l,l'} \delta (\omega ' - \omega ))$,

\begin{align}
\label{eq2}
\hat {H}_1 = &
\hbar {\sum\nolimits_{j = 1}^{N} {(\Delta
_{j} (t) + \delta \Delta _{j} (t))\hat {S}_{z}^j } }
\nonumber  \\&
+ \hbar \sum\nolimits_{l = 1}^2 {\int {(\omega - \omega _o )\hat {b}_l^ +
(\omega )\hat {b}_l (\omega )d\omega } }
\nonumber  \\&
+ i\hbar {\sum\nolimits_{j = 1}^{N } {[g_{j }
\hat {S}_ - ^{j } \hat {a}^ + - g_{j }^\ast \hat {S}_ + ^{j } \hat
{a}]} }
\nonumber  \\&
+ i\hbar \sum\nolimits_{l = 1}^2 {\int {\kappa _l (\omega )} [\hat {b}_l
(\omega )\hat {a}^ + - \hat {b}_l^ + (\omega )\hat {a}]} d\omega .
\end{align}

The first term in (\ref{eq2}) comprises the perturbation energies of atoms where $\Delta
_j (t)$ is a controlled frequency detuning of j-th atom $\Delta _j (t < \tau
) = \Delta _j $ and $\Delta _j (t > \tau ) = - \Delta _j $,  $\delta
\Delta _j (t)$ is its fluctuating frequency detuning determined by local
stochastic fields, $g_j $ is a photon-atom coupling constant in the QED cavity,
$S_{+}^j $ and $S_{ -}^j $ are the transition spin operators. Ensemble distributions
over the detunings $\Delta _j $ and $\delta \Delta _j (t)$ determine
inhomogeneous $\Delta_{in}$ and homogeneous $\gamma_{21}$ broadenings of the resonant atomic line. In the following we use Lorentzian shape for inhomogeneous broadening (IB) and typical anzatz for ensemble average over the fluctuating
detunings $\delta\Delta_j(t)$:

\begin{align}
\label{eq3}
&\sum\limits_{j = 1}^{N} {\vert g_{j } \vert ^2} \exp \{ - i\Delta _j (t
- t')\}\Phi _{j} (t,t')\equiv
\nonumber   \\  &
 N \vert \bar {g} \vert ^2\exp \{ -
(\Delta _{in} + \gamma _{21} )\vert t - t'\vert \},
\end{align}

\noindent
where $\Phi _{j} (t,t') = \exp \{ - i\delta \varphi _{j} (t,t')\}$,
$\delta \varphi _{j} (t,t') = \int_{t'}^t {dt} "\delta \Delta _{j}
(t")$, $\vert \bar {g} \vert ^2$ is quantity averaged over the atoms. Second term in (\ref{eq2}) contains frequency detunings of the $l$-th field modes. Third terms is the interaction energy of atoms with cavity mode. Fourth term
is an interaction energy of the cavity mode with the signal and bath modes characterized by the coupling constants $\kappa _l (\omega )$.

We note that $[\hat {H}_o ,\hat {H}_1 ] = 0$ so a total number of excitations in the atomic system and
the field modes is conserved during the quantum evolution, $\hat
{H}_o $ gives a contribution only to the common phase of
the wave function. H$_{1}$ determines a unitary operator $\hat {U}_1 (t) =
\exp \{ - i\hat {H}_1 t / \hbar \}$ causing the evolution of the
atomic and field systems with dynamical exchange and entanglement of the
excitations among them.  We assume that initially all atoms ($j =
1,2,...,N )$ stay on the ground state $\left| g \right\rangle _a = \left|
{g_1 ,g_2 ,...,g_{N} } \right\rangle $  and we launch a
signal multi-mode single photon fields prepared in the initial quantum state
$\left| {\psi _{in} (t)} \right\rangle _{ph} = \prod\nolimits_{k = 1}^M
{\hat {\psi }_k^ + (t - \tau_k )} \left| 0 \right\rangle $, $\hat {\psi }_k^ +
(t - \tau_k ) = \int_{ - \infty }^\infty {d\omega _k f_k (\omega _k )\exp \{ -
i\omega _k (t - \tau_k )\}\hat {b}_1^ + (\omega _k )} $; $f_k$ is a wave function in the frequency space
normalized for pure single photon state $\int_{ - \infty}^\infty {d\omega _k \vert f_k (\omega _k )\vert ^2 = 1} $, M is a number of modes, $\left| 0 \right\rangle $ is
a vacuum state of the field. k-th photon mode arrives in the circuit at time moment
$\tau_k $, time delays between the nearest photons are assumed to be large
enough $(\tau_k - \tau_{k - 1} ) \gg \delta t_k$, $\delta t_k$ is a temporal duration
of the k-th field mode. Additional bath field modes ($l=2$) are assumed to be in
the vacuum state $b_2 (\omega )\left| 0 \right\rangle = 0$.
Thus, the total initial state of the multi-mode light field and atoms is given by $\left| {\Psi _{in} (t)}
\right\rangle = \left| {\psi _{in} (t)} \right\rangle _{ph} \left| g
\right\rangle _a $.

Neglecting a population of excited atomic state and using the input and
output field formalism \cite{Milburn} we derive the following linearized system of
Heisenberg equations for the field operators and for the atomic operators in the rotating frame representation:

\begin{equation}
\label{eq4}
\textstyle{d \over {dt}}\hat {b}_l (\omega ) = - i(\omega - \omega _0 )\hat
{b}_l (\omega ) - \kappa _l (\omega )\hat {a},
\end{equation}

\begin{equation}
\label{eq5}
\textstyle{d \over {dt}}\hat {S}_ - ^{j }  = - g_{j}^\ast \hat {a} -
i[\Delta _{j } + \delta \Delta _{j} (t)]\hat {S}_ - ^{j} ,
\end{equation}

\begin{equation}
\label{eq6}
\textstyle{d \over {dt}}\hat {a} = {\sum\nolimits_{j =
1}^{N } {g_{j } } \hat {S}_ - ^{j } } - \textstyle{1 \over
2}(\gamma _1 + \gamma _2 )\hat {a} + \sqrt {\gamma _1 } \hat {b}_1 (t)
+ \sqrt {\gamma _2 } \hat {b}_2 (t),
\end{equation}

\noindent
where $\gamma _l = 2\pi \kappa _l^2 (\omega _o )$, ($l$=1,2) \cite{Milburn}. The input signal field containing M temporally separated photon modes
is given by
$\hat {b}_1 (t) = \sum\nolimits_{k = 1}^M {\hat {b}_{1,k} (t - \tau_k )}$ ,
where $\hat {b}_{l,k} (\tilde\tau_k ) = \textstyle{1 \over {\sqrt {2\pi }
}}\int_{ - \infty }^\infty {d\omega \exp \{ - i(\omega - \omega _0 )
\tilde\tau_k \} \hat {b}_l (\omega )}$ and obviously  we have $\hat {b}_2 (t) \left| 0 \right\rangle = 0$.

\emph{Quantum storage:} By taking into account  a formal
solution of Eq. (\ref{eq5})

\begin{align}
\label{eq7}
\hat {S}_ - ^{j } (t) = & \hat {S}_ - ^{j } (t_o )\Phi _{j} (t,t_o)\exp \{ - i\Delta _{j}
(t - t_o )\}
\nonumber  \\  &
 - g_{j }^\ast \int\limits_{t_o }^t {dt'}
 \Phi _{j} (t,t')
 \exp \{ - i\Delta _{j } (t -
t')\}\hat {a}(t'),
\end{align}

\noindent
we use the Laplace transformation for
$\hat {a}_L (p) = \int_{t_0 }^\infty {dte^{ - p(t - t_0 )}\hat {a}(t)} $ and
similarly for $\hat {b}_{1,L} (p))$ in (\ref{eq6}) that leads
$\hat {a}_L (p) = \sum\nolimits_{n = 1}^{4} \hat {a}_{n,L} (p) $, where

\begin{equation}
\label{eq8}
\hat {a}_{1,L} (p) = f(p)\hat {a}(t_o ),
\end{equation}

\begin{equation}
\label{eq9}
\hat {a}_{2,L} (p) = f(p)\sum\nolimits_{j = 1}^{N } {g_{j } } \hat
{S}_ - ^{j } (t_o )\Phi _{j}^{(s)} (p),
\end{equation}

\begin{equation}
\label{eq10}
\hat {a}_{3,L} (p) = f(p) \sqrt {\gamma _2 } \hat {b}_{2,L}(p),
\end{equation}

\begin{equation}
\label{eq11}
\hat {a}_{4,L} (p) = f(p)\sqrt {\gamma _1 } \hat {b}_{1,L} (p),
\end{equation}

\noindent
where $\Phi _{j}^{(s)} (p) = \int_{t_o }^\infty {dt \exp\{ -( p+i\Delta_j )(t - t_o )\}} \Phi _{j } (t,t_o )$,
$\hat {b}_{l,L} ( - i\omega ) = \sqrt {2\pi } \hat {b}_l (\omega)$,

\begin{equation}
\label{eq12}
f(p) = \left( {p + \frac{(\gamma _1 + \gamma _2 )}{2} + \frac{N_1 \vert
\bar {g} \vert ^2}{(p + \Delta _{in} + \gamma _{21} )}} \right)^{ - 1}.
\end{equation}

After inverse Laplace transformation, we find a solution $\hat {a}(t) =
\sum\nolimits_{n = 1}^4 {\hat {a}_n (t)} $, where four terms of the cavity
field $\hat {a}_n (t) = \textstyle{1 \over {2\pi }}\int_{\varepsilon -
i\infty }^{\varepsilon + i\infty } {dpe^{p(t - t_0 )}\hat {a}_{n,L} } (p)$
have different temporal and physical properties. The first field $\hat {a}_1 (t)$
is determined by the initial field $\sim \hat {a}_1 (t_0 )$ that disappears rapidly in the cavity
on time interval $(t - t_o ) > [\textstyle{1 \over 2}(\gamma _1 +
\gamma _2 ) + N_1 \vert \bar {g}_1 \vert ^2 / (\Delta _{in} + \gamma _{21}
)]^{ - 1}$. Second  field component $\hat {a}_2 (t)$  is excited due to the
interaction with  atomic coherence at $t=t_o$.  Third $\hat {a}_3 (t)$ and fourth $\hat {a}_4 (t)$ field components  are excited by the bath modes $\hat {b}_2 (\nu )$ and by the signal field $\hat {b}_1 (t)$. Due to the initial state
$\left| {\Psi _{in} (t)}\right\rangle$,
the field components  $\hat {a}_2 (t)$ and  $\hat {a}_3 (t)$ redetermine only the QED cavity vacuum without excitation of real photons. By taking into account the expectation values $ \langle \hat {b}_2^ + (\nu)\hat
{b}_2 (\nu) \rangle = \langle S_{ +}^j (t_o )S_{ -}^j (t_o ) \rangle = 0$
for initial state, we leave only the nonvanishing term for the atomic coherence at $t=\tau$ determined by the signal field

 \begin{align}
\label{eq13}
\hat {S}_ - ^{j } (\tau) =
 - g_{j }^\ast \int\limits_{t_o }^{\tau} {dt'}
 \Phi _{j} (\tau, t') \exp \{ - i\Delta _{j } (\tau -t')\}\hat {a}_4(t'),
\end{align}

By using (\ref{eq13}) and (3), we calculate a storage efficiency of the signal field
$Q_{ST}(\tau) = \bar {P}_{ee}(\tau) / \bar {n}_1 $ where $\bar {P}_{ee}(\tau) =
\sum\nolimits_{j = 1}^{N } { \langle  S_{ + }^j (\tau)S_{ - }^j (\tau) \rangle } $ is an
excited level population of atoms after the interaction with last M-th
signal field mode for $\tau > \tau_{M}+\delta t_M$
(we assume usual relation for temporal duration and spectral width of k-th mode
$\delta t_k \approx\delta\omega_k^{-1}$).
 Total number of photons
in the input signal field is $\bar {n}_1 =
\sum\nolimits_{k = 1}^M {\bar {n}_{1,k} } $,  $\bar {n}_{1,k} = \int_{ -
\infty }^\infty {dt \langle \hat {b}_{1,k}^ + (t)} \hat {b}_{1,k} (t)\rangle $ is initial number of photon in k-th temporal mode, $\langle...\rangle$ is a
quantum averaging over the initial state $\left| {\Psi _{in} (t)}
\right\rangle$.
Performing the algebraic calculations of $\bar {P}_{ee}(\tau) $, we find the quantum
efficiency of storage $Q_{ST} = (1 / \bar {n}_1 )\sum\nolimits_{k = 1}^n {Q_{ST,k} }
\bar {n}_{1,k} $ where the storage efficiency  of k-th mode is

\begin{align}
\label{eq14}
Q_{ST,k} = &
\int\limits_{ - \infty }^\infty {d\nu }
{\rm Z}(\nu ,\Delta _{tot} ,\Gamma _{tot} )
\frac{\left\langle {\hat {n}_{1,k} (\nu)}\right\rangle}{\bar {n}_{1,k}},
\end{align}

\noindent
where spectral function
\begin{align}
\label{eq15}
&{\rm Z}(\nu ,\Delta _{tot} ,\Gamma _{tot} ) =
\nonumber  \\  &
\frac{\Delta _{tot}^2 }{(\Delta
_{tot}^2 + \nu ^2)}\frac{4\gamma _1 \Gamma _{tot} }{\vert \gamma _1 +
\gamma _2  + \frac{\Gamma _{tot} }{(1 - i\nu / \Delta _{tot} )}- 2i\nu \vert ^2},
\end{align}

\noindent
$\Gamma_{tot}  = 2N \vert \bar {g} \vert ^2 / \Delta _{tot}$
is a photon absorption rate by N-atomic ensemble in unit spectral
domain within the IB line, $\Delta _{tot}=\Delta _{in} + \gamma_{21} $ is a total line width.
For relatively narrow spectral width $\delta \omega _k $
of the k-th signal
field
and weak atomic decoherence rate in comparison with
IB ($\delta \omega _{k}, \gamma _{21} \ll \Delta _{tot}$), we get from Eqs. (\ref{eq14}) and (15):

\begin{equation}
\label{eq16}
Q_{ST,k} = \frac{\gamma _1 }{(\gamma _1 + \gamma _2 )}\frac{4\Gamma_{tot} /
(\gamma _1 + \gamma _2 )}{[1 + \Gamma_{tot} / (\gamma _1 + \gamma _2 )]^2}.
\end{equation}

\noindent
Quantum efficiency $Q_{ST,k} $ reaches unity at \emph{optimal matching conditions} $\Gamma_{tot} / \gamma _1
= 1$ and $\gamma _2 / \gamma _1 \ll 1$ that provide a
perfect storage of each k-th signal mode (maximum number of the modes is limited by $M_{\max } \sim \textstyle{{\Delta _{in}}/{\gamma _{21}}} \gg 1)$. At the optimal matching, the each input k-th mode  entering in the QED cavity will completely transfer to the atoms since the IB atomic system absorbs each spectral component of the input mode $f_k(\omega_k)$ with the same optimal rate $\Gamma_{tot}= \gamma _1$ leading to relation $\hat a (t)=\sqrt {\gamma_1} \hat b_1 (t)$ for arbitrary temporal mode shape that means an absence of the reflection of any input signal field from the QED cavity. So the storage of the multi-mode field in IB atomic system will occur by one step procedure.

\emph{Multi-mode Quantum Memory:} In accordance with photon echo QM protocol \cite{Moiseev2001}, after complete absorption of the signal QM, we recover the rephasing atomic coherence by changing the frequency detunings of atoms at time moment
$t=\tau $. Inversion of the atomic detunings
$\Delta _j\to- \Delta _j$ can  be done by using a  Doppler effect \cite{Moiseev2001}, properties of local fields \cite{Moiseev2003}, or by direct changing a polarity of the external magnetic or electric fields \cite{Kraus2006, Alexander2006}. It is also possible to recover the atomic coherence with quite large efficiency by exploiting the atomic frequency comb structure of the IB line \cite{Afzelius2009}. At first, we demonstrate the QM  in QED cavity in most general way by using a Schr\"{o}dinger picture by taking into account the interaction with cavity mode and other field modes.

In spite of huge complexity of the compound
light-atoms system, here, we show that their quantum dynamics governed by
H$_{1}$ in (\ref{eq2}) can be perfectly reversed in time on our demand in a simple
robust way.
By transferring to the new field operators $\hat
{a} = - \hat {A}$ and $\hat {b}_1 (\omega _o - \Delta \omega ) = \hat {B}_l
(\omega _o + \Delta \omega )$  (with similar relations for the Hermit
conjugated operators),  we get a new Hamiltonian with an opposite sign
  in comparison with initial one $\hat {H}_1 ' = - \hat {H}_1 $ determining a temporally reversed evolution  $\hat {U}_2 [(t - \tau )] = \exp \{ - i\hat {H}_1 '(t - \tau ) /
\hbar \} \quad  = \exp \{i\hat {H}_1 (t - \tau ) / \hbar \}$. Ignoring weak
interaction with the bath modes and slow atomic decoherence, i.e. assuming
$\gamma _{21} \approx 0,\gamma _2 \approx 0$, we find that the quantum evolution $\hat {U}_2 $
recovers the initial quantum state of the multi-mode signal field and atoms at
$t = 2\tau $ due to unitary reversibility of the echo
signal emission making the echo field spectrum inverted relatively to the central
frequency $\omega _o $ in comparison with the original one.

Coming back to the real parameters of atomic decoherence rate $\gamma _{21}$ and cavity parameters  $\gamma _{1}$, $\gamma _{2}$,  we analyze below a retrieval of the echo field and QM efficiency for the multi-mode signal field (the field index $"e"$ is introduced to indicate the echo emission stage). By changing $\Phi _{j}^{(s)} (p)$ to
$\Phi _{j}^{(e)} (p) = \int_{\tau }^\infty {dt \exp\{ -( p-i\Delta_j )(t - \tau)\}} \Phi _{j } (t,\tau )$,
we find the Laplace image of the quantum echo field irradiated by the atomic coherence  $\hat {S}_ - ^{j } (\tau)$ in accordance with Eq. (9). We find the echo field in time domain picture $\hat {a}_{e} (t)$  after inverse Laplace transformation, calculation of all temporal integrals and summation over the atomic responses. By taking into account large IB in comparison with the atomic decoherence rate $\Delta_{in}\gg\gamma_{21}$, we find the  echo field irradiated in the QED cavity

\begin{align}
\label{eq17}
& \hat {a}_{e} (t) =
-\frac{\exp \{ - 2\gamma _{21}(t - \tau )\}}{\sqrt {\gamma _1 }}
{\int\limits_{ - \infty }^{ + \infty }
{\frac{d\nu }{\sqrt {2\pi } }}}
\nonumber  \\   &
\sum\nolimits_{k = 1}^M
{\rm Z}(\nu ,\Delta _{in} ,\Gamma _{in} )
\hat {b}_{1,k} (\nu )\exp \{i\nu (t + \tau _k - 2\tau )\},
\end{align}

\noindent
where $\Gamma _{in} = 2N\vert \bar {g}\vert ^2 / \Delta _{in}\approx \Gamma_{tot}$.
The total photon number operator of the echo field signal irradiated at time $t\gg 2\tau$ is
$\hat {n}_{e} = \int_{ - \infty }^\infty {d\nu } \hat {b}_{e}^ + (\nu)
\hat {b}_{e} (\nu)  = \sum\nolimits_{k = 1}^M {\hat
{n}_{e,k} }$,
where $\hat {n}_{e,k} = \gamma _1 \int_{  \tau }^\infty {dt'} \hat
{a}_{e,k}^ + (t')\hat {a}_{e,k} (t')$ relates to the $k-th$ field mode with average photon number
$\left\langle {\hat {n}_{e,k} } \right\rangle =
\exp \{ - 4\gamma _{21}(\tau - \tau _k )\}
Q_{ME,k}\bar {n}_{1,k}$ and

\begin{equation}
\label{eq18}
Q_{ME,k}=
\int\limits_{ - \infty }^{ + \infty } {d\nu} [{\rm
Z}(\nu,\Delta _{in} ,\Gamma _{in} )]^2
\frac{\left\langle {\hat {n}_{1,k} (\nu)}\right\rangle}{\bar {n}_{1,k}}.
\end{equation}

\noindent
We have assumed in (18) that $\gamma_{21}\delta t_k \ll 1$. A spectral function  $[{\rm Z}(\nu ,\Delta _{in} ,\Gamma _{in} )]^2$
filtering the echo spectrum demonstrates an influence of two similar steps of the light-atoms interaction in accordance with their temporal reversibility,
factor $\exp \{ - 4\gamma _{21} (\tau - \tau_k )\}$ is a result of the atomic decoherence on the QM efficiency during
the storage time $2(\tau - \tau_k )$ of the k-th mode.  Total quantum efficiency
$Q_{ME}$ of the multi-mode field retrieval is

\begin{equation}
\label{eq19}
Q_{ME}=\frac{1}{\bar {n}_{1}}
{\sum\nolimits_{k =1}^{M }}
\exp \{ - 4\gamma _{21}(\tau - \tau _k )\}
Q_{ME,k} \bar {n}_{1,k}.
\end{equation}

\begin{figure}
  \includegraphics[width=0.4\textwidth,height=0.3\textwidth]{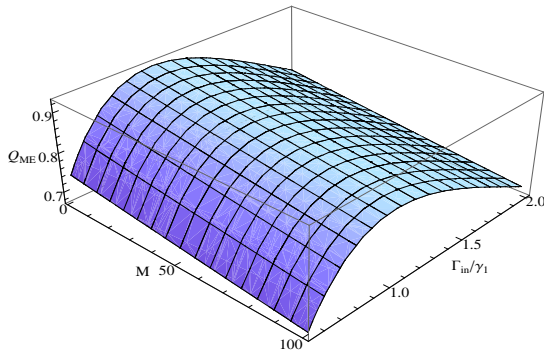}
  \caption{Quantum efficiency $Q_{ME}$ as a function of the mode number $M:=\{1,...,100\}$ and of the ratio $\Gamma_{in}/\gamma_1$
  for the atomic parameters: $\sqrt {N} |\bar g| =20$, $\Delta_{in}=10$, $\gamma_{21}=0.0001$ which are given in units of  spectral width $\delta\omega_k=\delta\omega=\delta t^{-1}$ of the input signal mode,   time delay between the nearest two temporal modes is $(\tau_{k+1}-\tau_k)/\delta t=5$, each mode contains $\bar {n}_{1,k}=1$ photon, $\gamma_2/\gamma_1=0.01$. }
  \label{Figure1}
\end{figure}

The quantum efficiency of the multi-mode field retrieval is depicted in
Fig  \ref{Figure1} in accordance with Eqs. (\ref{eq18})-(\ref{eq19}) for broad range of ratio $\Gamma_{in}/\gamma_1$ and reasonble parameters of the IB resonant atoms for Lorentzian spectral shape of each k-th temporal
field mode $ \left\langle {\hat {n}_{1,k} (\nu)}
\right\rangle =\langle\hat {b}_1^ + (\nu )\hat {b}_1 (\nu ) \rangle _k  =
 \textstyle{1 \over \pi } \bar {n}_{1,k}\delta \omega _{k} / (\delta \omega _{k}^2 + \nu
^2)$.

As seen in Fig  \ref{Figure1}, the QM efficiency for multi-mode field is close to unity at
optimal atomic and cavity parameters  $\Gamma_{in}/\gamma_1=1$. The QM efficiency
is equal unity in the theoretical limit $\gamma_1>\Delta_{in}\gg\delta\omega_k$ and $\gamma_2\ll\gamma_1$ if $\Gamma_{in}=\gamma_1$ leading to the following relation  for the optimal optical depth $\alpha L\sim \gamma_1 L/c$ (where c is a speed of light) that yields 100{\%} QM efficiency even for thin optical depth. To give an example, we assume $L=1$ mm and  $\gamma_1=10^{8}$ $s^{-1}$ that leads to the optimal optical depth $\alpha L\approx 3\cdot 10^{-4}$. Such small but optimal optical depth can be prepared by spectral tailoring of the original IB resonant line of rare-earth ions in the dielectric crystals \cite{Alexander2006,Tittel2010,Gisin2010}.

\emph{Conclusion:} We have found that an efficient multi-mode photon echo QM in QED cavity can be constructed at the optimal choice of the atomic and cavity parameters. Here, high QM efficiency can be realized even for the atomic system with thin optical depth determined by the matching condition for the atoms and cavity depending only on the spectral characteristics of the signal field (not on its temporal shape).
We stress a principle advantage of the proposed multi-mode photon echo QM in QED cavity
with respect to the QMs based on well-known EIT or early numerous
variants of photon echo QMs  \cite{Tittel2010} where 100 {\%} efficiency occurred only
for infinite optical depth of the coherent resonant atomic system ($\alpha L
\gg 1)$. So, using the QED cavity not only increases the optical depth via the well-known Purcell factor \cite{Purcell}, but makes its possible to realize an optimal condition for interaction between the external field and atoms. Thus, the predicted here possibility to get a highest efficiency
for the QM at finite values of the parameters easily achievable in experiment and its multi-mode character opens a door for practical applications.

\end{document}